# The opportunities and challenges of integrating population histories into genetic studies of diverse populations: a motivating example from Native Hawaiians

Charleston W.K. Chiang[1,2]

[1]Center for Genetic Epidemiology, Department of Preventive Medicine, Keck School of Medicine, University of Southern California
[2]Quantitative Computational Biology Section, Department of Biological Sciences, University of Southern California

Correspondence: charleston.chiang@med.usc.edu

**Abstract**
There is an urgent and well-recognized need to extend genetic studies to diverse populations, but several obstacles continue to be prohibitive, including (but not limited to) the difficulty of recruiting individuals from diverse populations in large numbers and the lack of representation in available genomic references. These obstacles notwithstanding, studying multiple diverse populations would provide informative, population-specific insights. Using Native Hawaiians as an example of an understudied population with a unique evolutionary history, I will argue that by developing key genomic resources and integrating evolutionary thinking into genetic epidemiology, we will have the opportunity to efficiently advance our knowledge of the genetic risk factors, ameliorate health disparity, and improve healthcare in this underserved population.

**Introduction**

Genome-Wide Association Studies (GWAS) have revealed the polygenic nature of human complex traits and diseases[1–3], but these successes are heavily biased towards European populations[4–6]. In order to truly personalize medicine for everyone, we need to combine an improved understanding of environmental and lifestyle risk factors with a better appreciation of the genetic etiology of complex diseases in geographically diverse, often underserved, populations. It remains a challenge to attain sample sizes from diverse populations comparable to existing European cohorts (>1 million individuals). Even when genetic data from understudied populations are included, they are often a small contributing part of a larger consortium such that any population-specific effects would likely be overshadowed. There is thus an urgent need to bring forth the benefits of genomic medicine to diverse populations through focused efforts. Whereas consortium-scale sample sizes are required to elucidate the polygenic nature and to detect individual variants with ever-decreasing effect sizes of a complex trait, the genetic causes of phenotypic differences among populations result from the distinct population history and unique interactions with the environment of the past or the present, which can be learned from detailed epidemiological and genetic data from moderately sized studies. For understudied populations, the focus is therefore both to transfer knowledge gained from large-scale Euro-centric studies, and to supplement our understanding with insights specific to the population at hand.

Genetic and phenotypic differences between populations can arise through two broad categories of evolutionary mechanisms: demographic events and natural selection. An example of demographic events is a population bottleneck. In a bottlenecked population, alleles with functional, deleterious consequences can by chance overcome the impact of negative selection[7] to reach higher frequencies and in turn explain a greater proportion of the heritability of a complex trait compared to alleles in a non-bottlenecked population[8–10]. An example of natural selection is through local adaptation to a variety of selective pressures such as climate, diet, UV exposures, or pathogens[11–13]. Alleles underlying adaptive traits will increase in frequency in the local population. But as the environment changed in modern societies, these adaptations could manifest as diseases and contribute to differences in risk between populations[14–16]. Leveraging these evolutionary events in practice has already identified population-enriched alleles disproportionately contributing to human complex traits in multiple populations around the globe[10,17–26]. These discovered alleles are oftentimes rare and difficult to map in large continental populations, but were found using only a moderately sized (by GWAS standards) cohort. Therefore, a better understanding of our evolutionary past will enable better designs and interpretations of genetic epidemiology studies, help explain the disparity in risks among populations today, and allow incorporation of evolutionary insights into our clinical practice[14]. However, these questions have not been systematically investigated in geographically diverse populations around the globe.

In this Perspective, we will use the Native Hawaiians as an example to illustrate the challenges and benefits to integrate evolutionary insights with genetic studies of diverse populations. Though they are one of the smallest ethnic minorities in the U.S., consisting of 1.2 million individuals and 0.4% of the U.S. census in 2010, Native Hawaiians and other Pacific Islanders (alone or in combination with other races) showed the second fastest rate of growth at 40% between 2000 to 2010. Compared to European- or Asian-Americans, Native Hawaiians display alarming rates of obesity, diabetes, cardiovascular diseases, cancers, and other related chronic health conditions[27–34]. Oftentimes their risks for diseases are elevated even after adjusting for BMI and other socioeconomic and lifestyle factors[28–31]. This suggests that in addition to environmental factors, systematic differences in the number, frequencies, or effect sizes of genetic risk alleles could help explain the risk disparity among populations. The history of Native Hawaiians exemplifies all major evolutionary mechanisms influencing the pattern of variations in humans – population size changes, adaptation, and recent admixture. We will describe the opportunities to leverage extensively characterized genetic history for understanding the Hawaiian-specific disease architecture, current challenges that inhibit large-scale and systematic genetic studies, and important considerations of conducting studies in this population. While we use Native Hawaiians as an example for motivation, the opportunities and challenges described here are generally applicable to all understudied populations around the globe.

**Demographic and admixture history of Native Hawaiians**

There is no detailed characterization of the demographic history of Native Hawaiians using genetic data. Based on archaeological findings, ancient DNA studies, and oral history, we know that ancient Austronesians originated from Taiwan and traveled through northern Philippines to arrive at the remote reaches of Oceania and Western Polynesia about 2,000 to 3,000 years ago[35–38]. These Austronesians settled in islands like Vanuatu, Tonga, and Samoa for nearly 1,000-2,000 years[39,40], where they coinhabited with the Papuans, the native people of Oceania. Today, Polynesian populations including the Native Hawaiians[41] have varying levels of an ancestry found predominantly in present-day Papuans[35,37,38]. The ancient Polynesians began long-range seafaring to the vast stretches of the Pacific around 200 B.C. to 700 A.D., arriving at Hawai'i between 900 A.D. to 1300 A.D.[39,42,43]. Initially there were frequent interactions among inhabitants of various Polynesian islands, including possible contact with people from as far as the Western Coast of South America[44]. The interactions ceased by the 1400s, perhaps due to the development of more complex sociopolitical structures on these islands, and the Native Hawaiians became relatively isolated until the European settlers arrived[39,40]. Records of Native Hawaiian population sizes pre-European contact are unreliable, but the effective population sizes ($N_e$) for Native Hawaiians are likely small throughout history since a genetically-estimated $N_e$ as recent as 1,000 years ago was reported to be ~1,000 for Melanesians and Samoans[45,46]. Thus, the demographic history of the Native Hawaiians is likely characterized by multiple founding events and persistent small sizes, which would permit rare alleles to drift to higher frequencies and contribute uniquely to the genetic architecture. Like previous examples from

Sardinia, Peru, and Samoa[19–21,25], a moderate-sized cohort of Native Hawaiians and other Polynesians could provide power to detect these population-specific associations.

Another important aspect of the Native Hawaiian demographic history is recent admixture. There were tales of early shipwrecks that brought Japanese, European, or South American sailors to the Hawaiian archipelago[39,44], but the largest wave of migrants occurred following Captain James Cook's arrival in Hawai'i in 1778. Immigrants and missionaries from Europe and Americas as well as laborers from China and East Asia arrived throughout the 19th and 20th centuries. African-ancestry individuals began arriving on the island in the 20th century, mostly as part of the military force[39]. Today, Native Hawaiians are the most likely U.S. census group to report having two or more components of ancestry[47], deriving major continental ancestry from the Polynesians, Europeans, and East Asians[48]. Variations of these continental ancestries would also partly explain risks of diseases in Native Hawaiians. For example, an individual's proportion of Polynesian ancestry has been shown to be associated with the risk of obesity, while both Polynesian and East Asian components contribute to the risk of T2D (**Figure 1**). Note that Polynesian ancestry here is better considered as the component that spread across Polynesia from the initial settlements in remote Oceania. This component itself may be a mixture of the ancient Austronesians that showed close affinity to the East Asian ancestry, as well as the component ancestry native to Melanesia and found predominantly in Papuans today[35,40]. Moreover, while the associations of disease risks with Polynesian ancestry suggest the presence of Polynesian-specific genetic risk factors, the associations are also likely to reflect any cultural or environmental non-genetic factors correlated with Polynesian ancestry (*e.g.* diet). Nevertheless, the admixed nature of Native Hawaiian genomes suggests that approaches such as admixture mapping[49,50] could identify regions of the genome disproportionately impacting the health of Native Hawaiians.

**Potential role of adaptation in shaping the genetic architecture**

Numerous adaptive events likely shaped the genetic architecture of complex traits in Native Hawaiians. The successful settlement of previously uninhabited Hawaiian archipelago likely involved adopting new subsistence strategies and overcoming famines, nutritional deficiencies, and higher tropical load of infections[40]. The encounter in the 18th century with Europeans and their pathogens deeply impacted the Native Hawaiians: historians have suggested that pathogens such as syphilis, gonorrhea, measles, whooping cough, mumps, cholera, or smallpox, among others, contributed up to an 80% decrease in census size in Hawai'i between 1780 to 1850 (ref.[39]). Both diets and pathogens are known evolutionary forces that shaped the genomes of human populations and contributed to phenotypic differences between populations today[11–13]. As such, adaptation, whether due to forces of nature or actions of the people, could also leave a lasting imprint on the health of Native Hawaiians. However, this hypothesis has not been systematically tested in Native Hawaiians or any Polynesian populations.

Native Hawaiians, and Polynesian populations in general, are more susceptible to metabolic diseases such as obesity and type-2 diabetes[21,28,29,40,48,51]. One intriguing explanation for this elevated susceptibility is the "Thrifty Gene Hypothesis," which stipulates that efficient energy storage during times of famine in the past provided an evolutionary advantage that is no longer consistent with the present-day diets. This hypothesis could explain the higher burden of metabolic diseases observed in Polynesian populations today given that individuals most capable of conserving energy likely were chosen to make the arduous trans-Pacific voyage and to adapt to new environment. However, there are questions whether the diversity of environments and genetic ancestries across the Pacific would all converge on the same manifestation of risk for metabolic syndromes[40], and the genetic support for the Thrifty Gene Hypothesis in other populations has been inconclusive[52,53]. Recent genomic data in Samoans identified the derived allele of rs373863828 in *CREBRF* to be associated strongly with increasing BMI and protection for T2D (ref.[21]). The allele is uniquely found among Pacific Islanders. In Samoans, its frequency is ~26% and it exists on an unexpectedly long haplotype, consistent with positive selection and the Thrifty Gene Hypothesis[21]. Nonetheless, this is a single locus that could be selected for its pleiotropic effect in Samoa. Its frequency is quite variable across Pacific Island populations[26,54–58], and the selection evidence was not replicated in a recent study of Native Hawaiians (although the sample size was small[26]). Given the more comprehensive European- and East Asian-centric GWAS data on BMI and T2D[59–62], the advancement in population genetic methods to detect selection across different time scales[63–66], and the emerging genomic data from large epidemiological cohorts from Polynesian populations[21,48], there is now an opportunity to systematically survey the genome for signature of adaptation, assess their modern-day health consequences, if any, and rigorously put the Thrifty Gene Hypothesis to test.

**Challenges in genomic studies of Native Hawaiians**

One deterrent to including Native Hawaiians in genomic studies is the underdevelopment or unavailability of genomic resources. For other continental populations, these resources have been abundant and publicly available, enabling large-scale collaborations and investigations. Development of these resources in Native Hawaiians or other Polynesian populations will similarly accelerate genetic research in these populations at large.

One sorely-needed resource is a catalog of genetic variation, akin to gnomAD which contains variation discovered from sequencing data of up to ~141,000 individuals[67]. This catalog has substantially improved clinicians' ability to interpret clinical sequencing data of severe and rare genetic diseases and to reach a genetic diagnosis. Though still dominated by genomic data from European individuals, gnomAD does include data from ~20,000 individuals of African ancestry, and similar catalogs are emerging from Asians as well[68–71]. However, Native Hawaiians or Polynesians in general are not yet represented in these catalogs. The publicly available sequencing data of Native Hawaiians is limited to data from a single individual in the Simons Genome Diversity Project[72]. (There are also ~28 individuals across Oceania in the Human Genome Diversity Panel[45].) Going forward, the sample size need not be large -- even several hundred individuals will allow one to detect nearly all common variations (with allele frequency > 1%) in the population. Since many of these variants will be Polynesian-specific and have not been observed elsewhere in the world, such a catalog will further improve physicians' ability to interpret variants of unknown significance in the clinical setting to directly benefit the Polynesian community[73].

To accelerate the discovery of genetic associations to diseases, we also need to improve Native Hawaiian representation in imputation reference panels. Genome-wide genotyping, followed by imputation of the unobserved genetic variation, is one of the most economical approaches to conduct genetic association studies. Publicly available imputation reference panels are constantly growing in size, allowing investigators to query rarer variations that are usually absent on genotyping arrays. Because of the lack of representation in imputation reference panels, the quality of imputation in Native Hawaiians lags significantly behind that of other ethnic minorities (**Figure 2**). In a proof-of-principle study, it was shown that rs373863828 in *CREBRF* is associated with a large effect on BMI and T2D in Native Hawaiians, but could not be imputed or discovered using publicly available imputation resources at the time, despite the study having sufficient statistical power to do so[26]. The lack of representation has thus contributed to the disparity in bringing genomic medicine to Native Hawaiians compared to other ethnic minorities in the United States.

Ultimately, larger cohorts will boost statistical power and undoubtedly enhance the insights we can garner, but large recruitments in indigenous communities such as the Native Hawaiians have been challenging. The population sizes of any indigenous population are already small, and past mistakes by researchers, such as the Havasupai diabetes study that misused genetic information from the indigenous community in unconsented studies[74], have also caused indigenous communities to lose faith in scientists. In a recent assessment of Pacific Islanders, over 65% of participants shared some reservation or reluctance about providing biospecimens for research, citing concerns due to spirituality, lack of knowledge of research, or invasion of privacy, among others[75]. With increasing awareness of these past mistakes, genome scientists should open dialogue with the community early and often, respect both community and individual consent, and *partner with indigenous communities*, rather than just enrolling them as participants[74,76,77].

**Discussion**

Population genetic theories predict that there will be unique genetic variants segregating in Native Hawaiians. Identifying these variants, particularly those exerting a disproportionate impact on the health of Native Hawaiians, will significantly improve healthcare practices and directly benefit this community. Though several challenges currently exist, the outlook for genetic research in Native Hawaiians and other diverse populations in general can be promising while requiring only moderate level of funding commitments. Whole genome sequencing of only 150-200 Native Hawaiian individuals would already allow better imputation of Native Hawaiian individuals in a genetic study and accelerate the discovery of population-specific alleles of large effects[26,78]. The generation and aggregation of WGS data from multiple Polynesian populations will also provide the catalog of genetic variation currently lacking in Polynesian populations, make an immediate impact in the clinical care of Polynesian populations, and accelerate future large-scale genomic research in these populations. This roadmap can be achieved by pooling resources from a handful of research labs, and the cost can potentially be lowered by deploying low-coverage sequencing[19,70] as a first step. These are realistic outlooks over the next five years.

Although well-intentioned from a healthcare standpoint, extreme caution must also be taken to interpret and contextualize research findings through partnership with the Native Hawaiian community. For example, even though the quantification of components of genetic ancestries in Native Hawaiians is a necessary first step to dissect population-specific genetic risk factors, it should not supplant current approaches, such as through self-identification or genealogical records, to define community membership. First, estimated ancestry proportions are not without errors and can be sensitive to the choice of variants analyzed or reference panels used, particularly for a population of complex admixture history[79]. Moreover, there is a conceptual difference between genetic ancestry and genealogical ancestry. That is, an individual may not inherit any genetic material from a genealogical ancestor[80], but that should not detract from the individual's cultural identity or heritage. Furthermore, definitions of genetic ancestry can evolve as we develop more advance methods or integrate with different sources of data. Ultimately, the discrete nature of our models for genetic ancestry may be an arbitrary construct out of convenience, with roots in biological taxonomy, the bifurcating notion of the tree of life, and the limited geographical samplings of human populations.

Past exploitation of genetic data of indigenous populations has been attributed to the ignorance of tribal customs and regulations, or blatant patriarchal attitudes towards indigenous populations, among others[74,76,77]. The enormous amount of profit generated by biomedical advancements and pharmaceutical drugs have rarely trickled down to indigenous communities in the form of access to healthcare or reinvestment for infrastructures, and in turn harms the interest of the community[81]. Collectively, these factors brood mistrust between underprivileged communities and scientists. While pharmaceutical or biotech companies will be positioned to directly benefit indigenous communities with proceeds distributions or profit sharing, academic researchers will be better positioned to tailor their engagement to the unique circumstances of each community due to their long-term individualized interactions. That is, individual researchers should help build research capacity, actively engage in outreach and education, act in stewardship of indigenous data, and learn (and earn) to be an ally to the community.

**Acknowledgement**

I would like to thank John Novembre, Vivian U, and members of the Native Hawaiian Community Advisory Board at University of Hawaiʻi Cancer Center for their critical comments on earlier versions of this manuscript. I would also like to thank Xin Sheng, Victor Hom, and Bryan L. Dinh for assistance with imputation using the TOPMed reference panel. Computation for this work was supported by the Center for Advanced Research Computing (CARC) at the University of Southern California (https://carc.usc.edu).

# Figures

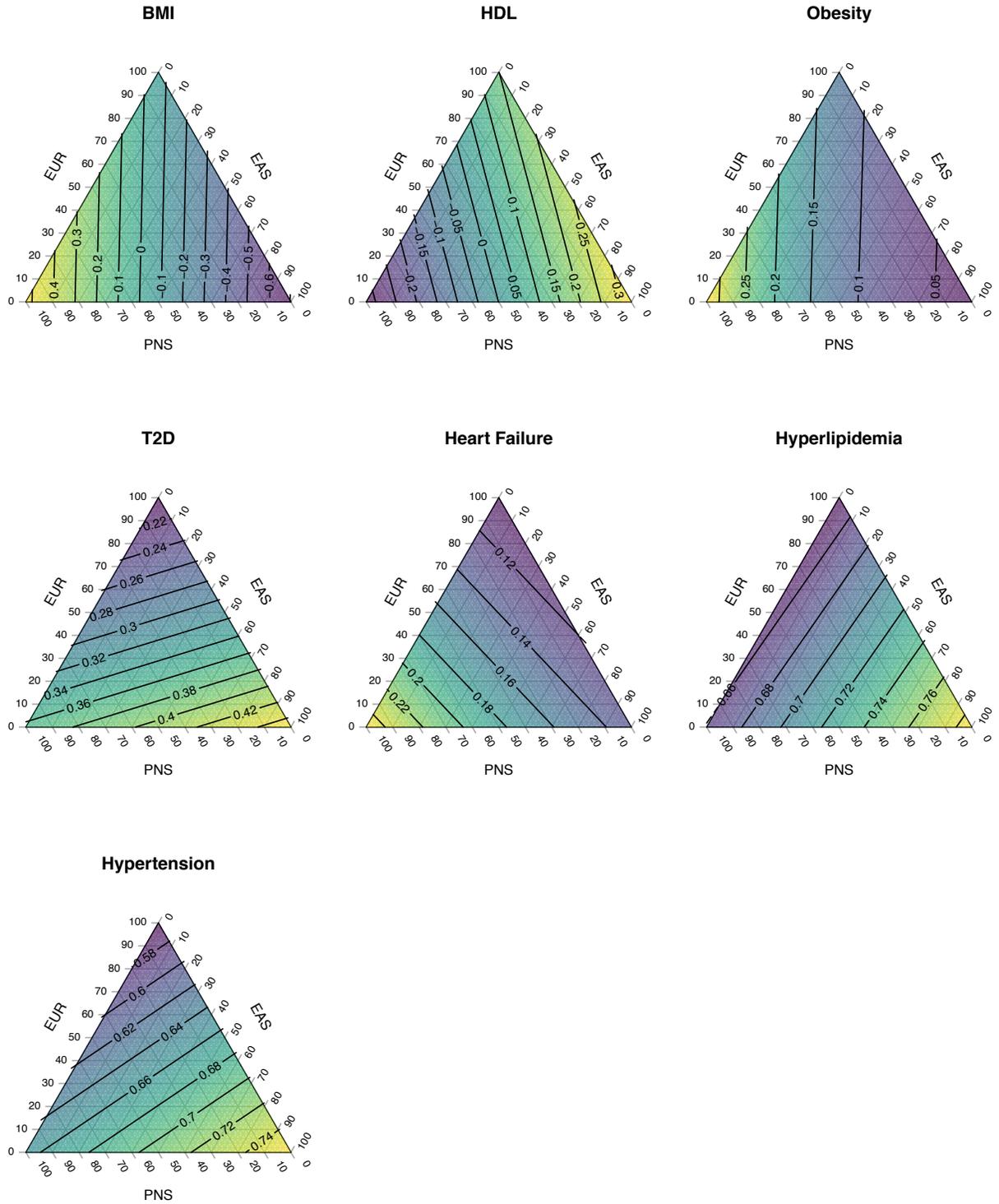

**Figure 1: Impact of ancestry components on complex traits and disease risks in Native Hawaiians.** For each of the seven traits and diseases, the estimated trait values (in units of standard deviations for quantitative traits BMI and HDL) or probability of being affected (for remaining dichotomous phenotypes) were calculated using linear models of disease risk as function of ancestry components from ref. [48]. Fitted values were interpolated across all possible combinations of ancestries and shown with contour lines. For simplicity only the three major ancestry components for Native Hawaiians are modeled. For dichotomous traits we assumed fixed values for the following covariates: age = 50, BMI = 30 (except for obesity), sex = male, and education level = college graduates. EUR, EAS, and PNS denote European, East Asian, and Polynesian ancestries, respectively.

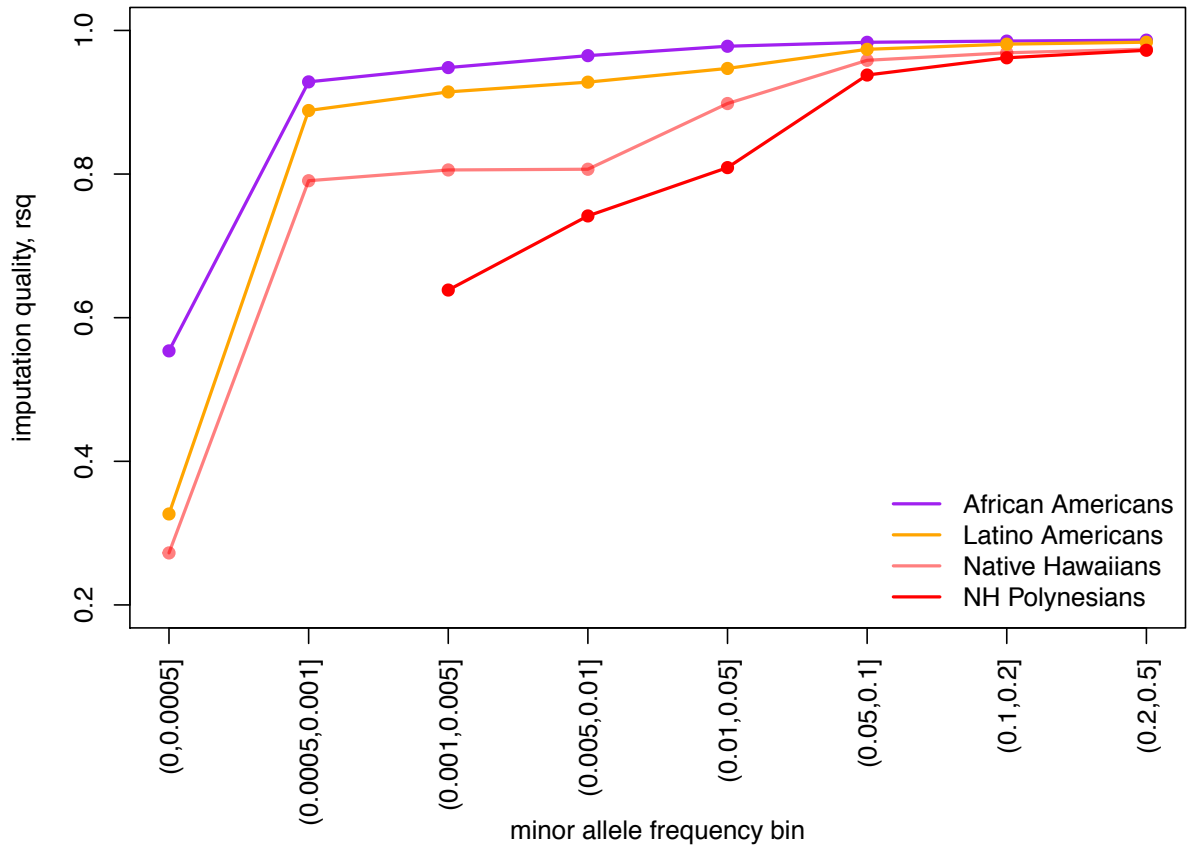

**Figure 2: Relatively poor imputation quality for Native Hawaiians due to underrepresentation in imputation reference panels.** We imputed 5,325 African Americans, 2,838 Latino Americans, and 3,940 Native Hawaiians from the Multiethnic Cohort[82] using freeze 8 of the TOPMED imputation server[83] (imputed in July 2020). Each population was genotyped on the MEGA array and subjected to the same QC filters. The mean imputation quality measured by $R^2$ showed that Native Hawaiian individuals are imputed more poorly than other U.S. ethnic minority populations, particularly for variants with minor allele frequency < 5%. The disparity is even stronger when focusing on only the 178 Native Hawaiians with estimated Polynesian ancestry > 90% (NH Polynesians)[26].